\documentclass[11pt]{article}
\usepackage[blocks]{authblk}
\usepackage{amsmath,amssymb,graphicx,url}
\usepackage{color}

\date{\today}

\title{A better conditioned Domain Wall Operator}

\author{H.\ Neff, Luzernerstrasse 43, 6330 Cham, Switzerland \footnote{Email: hartmutneff@aol.com}}

\begin{document}

\maketitle

\begin{abstract}
  A variation of the Domain Wall operator with an additional parameter
  $\alpha$ will be introduced. The conditioning of the new Domain Wall
  operator depends on $\alpha$, whereas the corresponding 4D
  propagator does not. The new and the conventional Domain Wall
  operator agree for $\alpha = 1$. By tuning $\alpha$, speed ups of
  the linear system solvers of around 20$\%$ could be achieved.
\end{abstract}

\section{Introduction}

A variation of the Domain Wall operator is suggested here. It
introduces a parameter $\alpha$ that appears only as a global factor
in the 4D matrix elements. Therefore, this generalization is simple in
structure and the Domain Wall formalism and the reduction to the 4D
Overlap formalism can be used almost unchanged. Details about the
Domain Wall and the Overlap formalism and how they can be translated
into each other can be found here \cite{Nielsen:1980rz,
  Nielsen:1981xu, Kaplan:1992bt, Callan:1984sa, Shamir:1993zy,
  Furman:1994ky, Borici:1999zw, Chiu:2002ir, Narayanan:1992wx,
  Narayanan:1993ss, Narayanan:1993sk, Neuberger:1997bg,
  Neuberger:1997fp,Kikukawa:1999sy, Edwards:2000qv, Brower:1997ha,
  Brower:2012vk}. As a reference for notation and for the sake of
completeness the standard 5D to 4D reduction will be rederived in
appendix \ref{sec:appa} and \ref{sec:appb}.

\section{The better conditioned Domain Wall operator}

The new Domain Wall operator introduces an additional parameter $\alpha$,
\begin{eqnarray}\label{eq:mobius1}
&D_{\alpha}(m) =&\nonumber\\ 
 &\hskip-0.4cm\left[\hskip-0.2cm\begin{array}{ccccc}
 D_{1+} (P_- + \alpha P_+ )  & \alpha  D_{1-} P_- &   0  &  \cdots&    -mD_{1-} P_+ \\
 \alpha D_{2-} P_+  &   \alpha D_{2+}  &   \alpha D_{2-} P_-  &  \cdots&   0  \\
 0  &   \alpha D_{3-} P_+  &  \alpha  D_{3+}  &   \cdots&  \cdots  \\
   \cdots&   \cdots&   \cdots&   \cdots&   \alpha  D_{(L_s-1)-} P_- \\
 -mD_{L_s-} P_-  &   0  &   \cdots & \alpha D_{L_s-} P_+ &  D_{L_s+} (P_+ + \alpha P_- ) \\
\end{array} \hskip-0.2cm \right] &
\end{eqnarray}
with
\begin{eqnarray}
&D_{i+} = b_i D_w +1 , \;\;\; D_{i-} = c_i D_w -1 \label{eq:coeff} ,& \\
&P_+ = \frac{1}{2} ( 1 + \gamma_5) , \;\;\; P_- = \frac{1}{2} ( 1 - \gamma_5) .&
\end{eqnarray}
$D_w$ denotes the Wilson Dirac matrix
\begin{eqnarray}\label{eq:wilson}
D_w (M_5)\hskip-0.05cm = \hskip-0.05cm (4+M_5) \delta_{x,y} -\hskip-0.1cm \frac{1}{2}  \bigl[  (1 \hskip-0.05cm   - \hskip-0.05cm   \gamma_\mu) 
 U_\mu(x) \delta_{x+\mu,y}  
\hskip-0.05cm  +  \hskip-0.05cm  (1  \hskip-0.05cm  +  \hskip-0.05cm   \gamma_\mu) U_\mu^\dagger(y) \delta_{x,y+\mu}  \bigr] .
\end{eqnarray}
Multiplying eq.(\ref{eq:mobius1}) from the right with $P$, (see eq.(\ref{eqP})), leads to

\begin{eqnarray} 
&D_{\alpha} P = D_{\alpha} 
\left[\begin{array}{ccccc}
 P_- &   P_+  &   0 &  \cdots&   0  \\
 0 &   P_-  &   P_+ &  \cdots&   0 \\
 0 &   0  &   P_- &  \cdots&    \cdots   \\
  \cdots&   \cdots&   \cdots&   \cdots&    P_+ \\
  P_+ &   0  & \cdots&    0 &   P_- \\
  \end{array} \right] & \\
&=\gamma_5  \left[\begin{array}{ccccc}
      Q_{1-} c_-  & \alpha  Q_{1+} & 0 & \cdots& 0  \\
      0 &  \alpha Q_{2-} & \alpha  Q_{2+}&  \cdots& 0  \\
      0 &  0& \alpha Q_{3-} &  \cdots&   \cdots \\
    \cdots&   \cdots&   \cdots&   \cdots&    \alpha  Q_{L_{s-1}+}  \\
      Q_{L_s+} c_+ & 0  & \cdots  & 0& \alpha Q_{L_s-}   \\
  \end{array} \right] &\\
& =D_{1} P  \left[\begin{array}{ccccc}
      1  &   0 & 0 & \cdots& 0  \\
      0 &   \alpha &   0&  \cdots& 0  \\
      0 &  0&  \alpha &  \cdots&   \cdots \\
    \cdots&   \cdots&   \cdots&   \cdots&      0 \\
      0& 0  & \cdots  & 0&  \alpha   \\
  \end{array} \right] \equiv D_{1} P A \; .&
  \end{eqnarray}
To find the 4D propagator, eq.(\ref{tosolve}) has to be solved,
  \begin{eqnarray}
      D_{1}(m) P  \vec{y} =D_{1}(1) P \vec{b}  ,
    \end{eqnarray}
  with source $\vec{b}$ and 4D propagator $y_1$.
  The independence of the 4D propagator from $\alpha$ follows directly,
    \begin{eqnarray}\label{tosolve1}
     &D_{1}(1) P \vec{b} = D_{1}(m) P  \vec{y} = D_{\alpha}(m) P A^{-1} \vec{y} = D_{\alpha}(m) P  \vec{z} \;,&
    \end{eqnarray}
with $A \vec{z} = \vec{y}$ and therefore $z_1 = y_1$.

\section{Results}

In this section, the $\alpha$ dependence of the conditioning of
$D_{\alpha}$ will be presented. The computations were done on 3 MILC
gauge fields of size $16^3 \times 32$, downloaded at NERSC.  The
conjugate gradient method on the normal equation was used to solve
eq.(\ref{tosolve1}).

The red black preconditioned version of $D_{\alpha}$ was used in the form,
\begin{eqnarray}\label{redblack}
  D_{bb} = 1_{bb} - I^{-1}_{bb} D_{br} I^{-1}_{rr}D_{rb} \; .
\end{eqnarray}
This version of red black preconditioning allows for an efficient use
of the Zolotarev approximation to the sign function. This is contrary
to what has been said in \cite{Brower:2012vk}, where we used the matrix,
\begin{eqnarray}\label{redblack1}
  D_{bb} = I_{bb} -  D_{br} I^{-1}_{rr}D_{rb} \; ,
\end{eqnarray}
instead. This is due to the fact that the rows of eq.(\ref{redblack1})
with large Zolotarev coefficients cause the convergence to slow
down. This behaviour can be improved by scaling all rows that contain
a Zolotarev coefficient larger than one with a factor equal to
the inverse of the Zolotarev coefficient. This can be seen as a
preconditioning from the left. But the even better method is to take
eq.(\ref{redblack}) where the preconditioning from the left cancels
out and where the weighting of the rows is done automatically.

The same behaviour can be observed for M\"obius coefficients $b_i$ and
$c_i$ larger than one.

The convergence of the linear system solver depends on the order of
the Zolotarev coefficients. The best performance was found, with an
ordering in a concave fashion, i.e.\ the smallest coefficients at
$i_5=1$ and $i_5 = L_s$ and the largest in the centre at $i_5 =
L_s/2$. Only then the parameter alpha improved the rate of
convergence. Zolotarev together with the parameter alpha should result
in a 2 to 3 times faster performance.

Let $n_i(\alpha)$ be the number of iterations for the residual to be
of the order of $O(-8)$, where $i$ runs over the color and Dirac
source indices and over the three gauge fields. The graphs in this
section show the relative count $n_i(\alpha)/n_i(\alpha=1)$, together
with the standard deviation, for a series of $\alpha$ values.

For the remaining parameters, the quark mass, the $5^{th}$ dimension
$L_s$ and the M\"obius coefficients $b_i$ and $c_i$, the optimal alpha
values and speed ups are summarised in the following table.
\begin{center}
  \begin{tabular}{c|c|c |c|c }\label{tab1}
    Mass & $L_s$ & $b_i$, $c_i$ & Best $\alpha$ & Speed Up \\ \hline
    0.06 & 4 & 1, 1 & 0.55 & 25\% \\ \hline
    0.06 & 6 & 1, 1 & 0.55 & 24\% \\ \hline
    0.06 & 8 & 1, 1 & 0.55 & 22\% \\ \hline
    0.06 & 10 & 1, 1 & 0.6 & 19\% \\ \hline
    0.06 & 12 & 1, 1 & 0.6 & 17\% \\ \hline
    0.01 & 8 & 1, 1 & 0.55 & 23\% \\ \hline
    0.06 & 8 & 1.7, 0.7 & 0.6 & 20\% \\ \hline
    0.06 & 10 & Zolotarev & 0.4 & 17\% \\ 
  \end{tabular}
\end{center}

\paragraph{Acknowledgment:}
I thank Richard Brower and Kostas Orginos for discussions and Tony
Kennedy for discussions and the code to compute Zolotarev
coefficients.

\begin{figure}
    \begin{center}
  \includegraphics[]{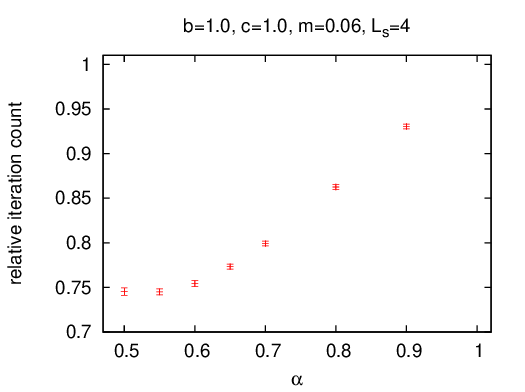} 
\end{center}
    \caption{\label{fig1}Relative iteration count $n_i(\alpha)/n_i(\alpha=1)$ for 3 gauge fields of size $16^3 \times 32$.}
\end{figure}

\begin{figure}
    \begin{center}
  \includegraphics[]{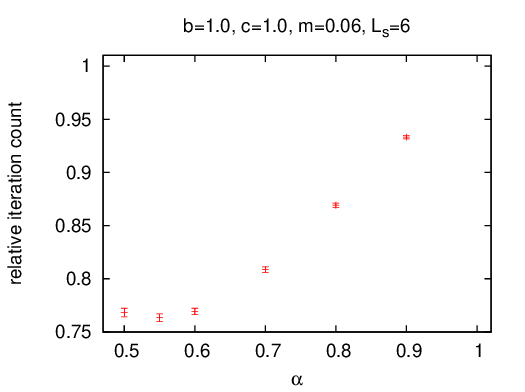} 
\end{center}
    \caption{\label{fig2}Relative iteration count $n_i(\alpha)/n_i(\alpha=1)$ for 3 gauge fields of size $16^3 \times 32$.}
\end{figure}

\begin{figure}
    \begin{center}
  \includegraphics[]{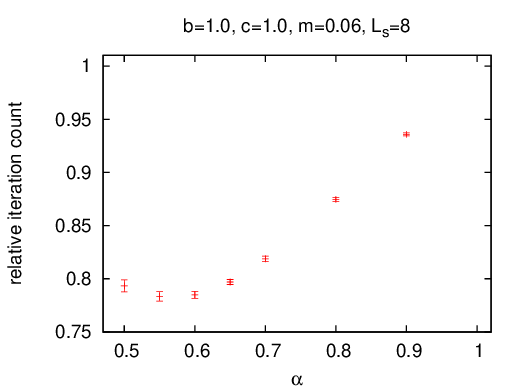} 
\end{center}
    \caption{\label{fig3}Relative iteration count $n_i(\alpha)/n_i(\alpha=1)$ for 3 gauge fields of size $16^3 \times 32$.}
\end{figure}

\begin{figure}
    \begin{center}
  \includegraphics[]{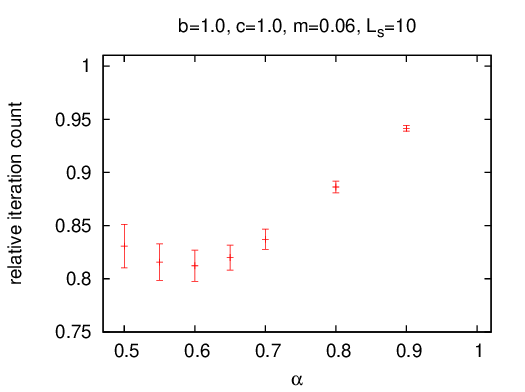} 
\end{center}
    \caption{\label{fig4}Relative iteration count $n_i(\alpha)/n_i(\alpha=1)$ for 3 gauge fields of size $16^3 \times 32$.}
\end{figure}

\begin{figure}
    \begin{center}
  \includegraphics[]{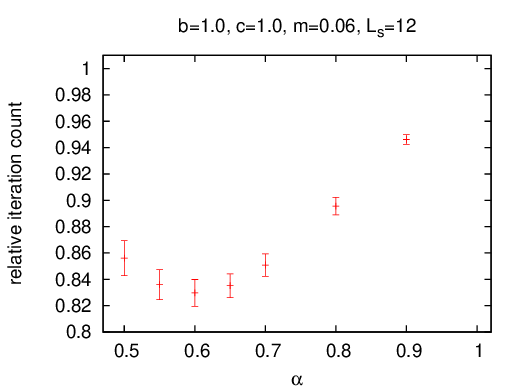} 
\end{center}
    \caption{\label{fig5}Relative iteration count $n_i(\alpha)/n_i(\alpha=1)$ for 3 gauge fields of size $16^3 \times 32$.}
\end{figure}

\begin{figure}
    \begin{center}
  \includegraphics[]{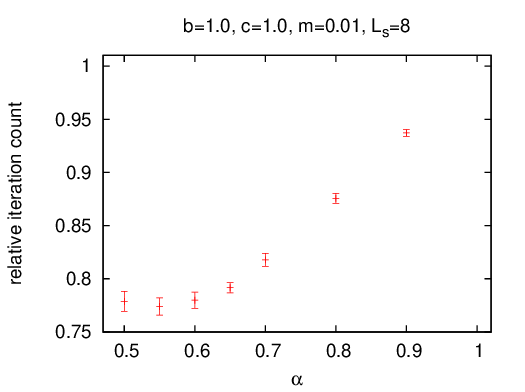} 
\end{center}
    \caption{\label{fig6}Relative iteration count $n_i(\alpha)/n_i(\alpha=1)$ for 3 gauge fields of size $16^3 \times 32$.}
\end{figure}

\begin{figure}
    \begin{center}
  \includegraphics[]{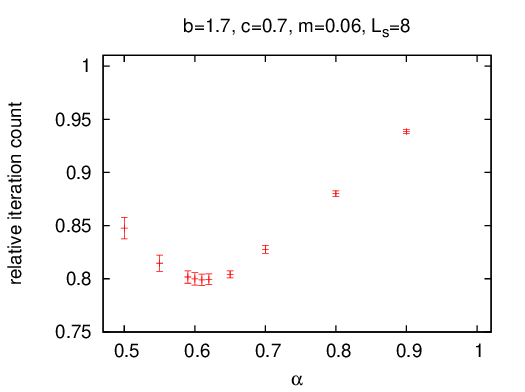} 
\end{center}
    \caption{\label{fig7}Relative iteration count $n_i(\alpha)/n_i(\alpha=1)$ for 3 gauge fields of size $16^3 \times 32$.}
\end{figure}

\begin{figure}
    \begin{center}
  \includegraphics[]{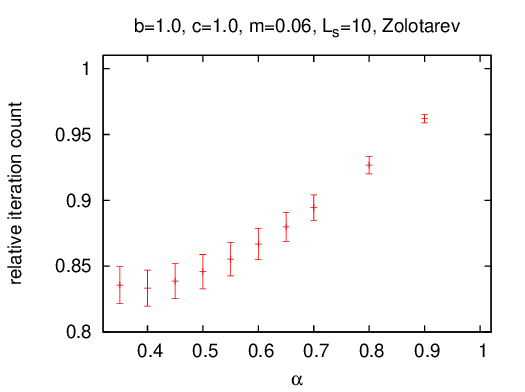} 
\end{center}
    \caption{\label{fig8}Relative iteration count $n_i(\alpha)/n_i(\alpha=1)$ for 3 gauge fields of size $16^3 \times 32$, with $b_i=c_i$.}
\end{figure}
\newpage

\bibliography{bettercondDW}

\newpage
\appendix
\section{Domain Wall to Overlap transformation}
\label{sec:appa}
To keep notation simple, we perform the transformation with $L_s=4$
sites in the 5th dimension. A generalisation to arbitrary $L_s$ is
straightforward.

  The Domain Wall to Overlap transformation reads,
\begin{eqnarray}\label{eq:4d-5d}
  L   D_{DW}(m) R  =  F  D^5_{OV}(m).
\end{eqnarray}
The transformation matrices take the form (for $L_s$ sites in the 5th dimension),
\begin{eqnarray}
F  = L   D_{DW}(1)  R ,
\end{eqnarray}
and
\begin{eqnarray}
L \hskip-0.05cm = \hskip-0.05cm  L_1 L_2 \hskip-0.05cm  = \hskip-0.05cm 
\left[\begin{array}{cccc}
1 & S_1   & S_1 S_2    & S_1 S_2 S_3   \\
0 & 1        & S_2           & S_2 S_3   \\
0 & 0        & 1                & S_3   \\
0 & 0        & 0                & 1 
\end{array} \right] \hskip-0.1cm
\left[\begin{array}{cccc}
Q_{1-}^{-1}  & 0  & 0 & 0  \\
0 & Q_{2-}^{-1}    & 0 & 0  \\
0 & 0  & Q_{3-}^{-1}   & 0  \\
0 & 0  & 0 & Q_{4-}^{-1}   \\
\end{array} \right]\gamma_5  , \nonumber
\end{eqnarray}
\begin{eqnarray} \label{eqP}\nonumber
R = P R_1 =
\left[\begin{array}{cccc}
 P_- &   P_+  &   0 &   0  \\
 0 &   P_-  &   P_+ &   0  \\
 0 &   0  &   P_- &   P_+  \\
 P_+ &   0  &   0 &   P_- \\
\end{array} \right] 
\left[\begin{array}{cccc}
  -1 &   0  &   0 &   0  \\
  -  S_2 S_3 S_4 \, c_+ &   1  &   0 &   0  \\
  -  S_3 S_4 \, c_+ &   0  &   1 &   0  \\
  -  S_4 \, c_+ &   0  &   0 &   1 \\
\end{array} \right],
\end{eqnarray}
\begin{eqnarray}
&& D^5_{OV}(m)= \left[\begin{array}{cccc}
D^4_{OV}(m) & 0  & 0 & 0  \\
0 & 1  & 0 & 0  \\
0 & 0  & 1 & 0  \\
0 & 0  & 0 & 1 \\
\end{array} \right].
\end{eqnarray}
The matrix entries are defined as follows,
\begin{eqnarray}
 & Q_{i+} = \gamma_5 D_w (b_i P_+ + c_i P_-) +1,\;\;\; Q_{i-} = \gamma_5 D_w (b_i P_- + c_i P_+) -1,& \nonumber\\
& S_i = T_i^{-1} = - Q_{i-}^{-1} Q_{i+} , &\nonumber\\
& c_+ = P_+ - m P_- ,\;\;\; 
 c_- = P_- - m P_+ . &
\end{eqnarray}
$T_i^{-1}$ is called the transfer matrix.

The matrix multiplications will be performed in the following order,
\begin{eqnarray}
L_1 L_2 D_{DW}(m) P R_1 = L_1 L_2 M_1 R_1 =  L_1 M_2 R_1 = L_1 M_3 = M_4 .
\end{eqnarray}
Step 1:
  \begin{eqnarray}
M_1  = D_{DW}(m) P 
   =\gamma_5  \left[\begin{array}{cccc}
      Q_{1-} c_-  &   Q_{1+} & 0 & 0  \\
      0 &   Q_{2-} &   Q_{2+} & 0  \\
      0 & 0  &  Q_{3-} &   Q_{3+}  \\
      Q_{4+} c_+ & 0  & 0 &  Q_{4-}   \\
    \end{array} \right],
  \end{eqnarray}
  with
  \begin{eqnarray}
   Q_{i-} & =&  \gamma_5 (D_{i+} P_- + D_{i-} P_+) \nonumber\\
    &=& \gamma_5( D_w (b_i P_- + c_i P_+) +P_- - P_+) \nonumber\\
    &=& \gamma_5 D_w(b_i P_- + c_i P_+) - 1,\\
    Q_{i+} &=& \gamma_5(D_{i+} P_+ + D_{i-} P_-)\nonumber\\
    & =& \gamma_5  ( D_w (b_i P_+ + c_i P_-) +P_+ - P_-)\nonumber\\
    &=& \gamma_5 D_w(b_i P_+ + c_i P_-) + 1 .
  \end{eqnarray}
Step 2:
  \begin{eqnarray}
  M_2 =  L_2 M_1 =
    \left[\begin{array}{cccc}
      c_-  & -  S_1 & 0 & 0  \\
      0 &  1  & -  S_2 & 0  \\
      0 & 0  & 1  & -    S_3 \\
      -S_4 c_+ & 0  & 0 & 1   \\
    \end{array} \right].
  \end{eqnarray}
Step 3:
  \begin{eqnarray}
  M_3 =  M_2 R_1=
    \left[\begin{array}{cccc}
      -c_- + S_1 S_2 S_3 S_4 c_+  & -  S_1 & 0 & 0  \\
      0 &  1  & - S_2 & 0  \\
      0 & 0  & 1  & -  S_3 \\
      0 & 0  & 0 & 1   \\
    \end{array} \right].
  \end{eqnarray}
Step 4:
  \begin{eqnarray}
  L   D_{DW}(m)  R= M_4 = L_1 M_3 =
    \left[\begin{array}{cccc}
      -c_- + S_1 S_2 S_3 S_4 c_+  & 0 & 0 & 0  \\
      0 &  1  & 0 & 0  \\
      0 & 0  & 1  & 0\\
      0 & 0  & 0 & 1   \\
    \end{array} \right].
  \end{eqnarray}
This leads to,
  \begin{eqnarray} 
 F = L D_{DW}(1) R =
    \left[\begin{array}{cccc}
      (1 + S_1 S_2 S_3 S_4 ) \gamma_5 & 0 & 0 & 0  \\
      0 &  1   & 0 & 0  \\
      0 & 0  & 1   & 0\\
      0 & 0  & 0 & 1    \\
    \end{array} \right].
  \end{eqnarray}
To make notation simpler, we define $S=S_1 S_2 S_3 S_4 $. The 5D Overlap Operator takes the form,
  \begin{eqnarray}\label{startprop}
 D^5_{OV}(m) = F^{-1} M_4  = 
    \left[\begin{array}{cccc}
      \gamma_5 (1 + S )^{-1} (-c_- + S c_+)  & 0 & 0 & 0  \\
      0 &  1  & 0 & 0  \\
      0 & 0  & 1   & 0\\
      0 & 0  & 0 & 1    \\
    \end{array} \right].
  \end{eqnarray}
It follows for the $(11)$ element,
\begin{eqnarray}
  D^5_{OV}(m)_{11} &=& \frac{1}{2}\gamma_5 (1 + S)^{-1}   \left( m + m \gamma_5 -1 + \gamma_5 + S (1+\gamma_5 -m +m\gamma_5)\right)\nonumber\\
&=&  \frac{1}{2}  \gamma_5(1 + S )^{-1}   \left( (1+m) (S+1) \gamma_5 +  (1-m) (S - 1)\right)\nonumber\\
&=&  \frac{1}{2}  \left( (1+m)   +  (1-m) \gamma_5\frac{ (S - 1)}{(S+1)}\right).
\end{eqnarray}
Hence eq.(\ref{startprop}) takes the form, 
  \begin{eqnarray}
 D^5_{OV}(m) =
    \left[\begin{array}{cccc}
      \frac{1}{2}  \left( (1+m)  +  (1-m) \gamma_5\frac{ (S - 1)}{(S+1)}\right) & 0 & 0 & 0  \\
      0 &  1  & 0 & 0  \\
      0 & 0  & 1  & 0\\
      0 & 0  & 0 & 1  \\
    \end{array} \right].
  \end{eqnarray}
  The matrix that acts as the variable for the polar decomposition can be found by setting,
    \begin{eqnarray}\label{polar}
    \frac{ (S-1)}{(S+1)} = \frac{(1-1/S)}{(1+1/S)}=
    \frac{\Pi_{i=1}^4(1+a_iX_i) -
      \Pi_{i=1}^4(1-a_iX_i)}{\Pi_{i=1}^4(1+a_iX_i) + \Pi_{i=1}^4(1-a_iX_i)},
    \end{eqnarray}
     and therefore
      \begin{eqnarray}\label{polar1}
         \frac{1}{S} = \frac{1}{S_1 S_2 S_3 S_4} =
        \frac{(1-a_1X_1(1-a_2X_2)(1-a_3X_3)(1-a_4X_4)}{(1+a_1X_1)(1+a_2X_2)(1+a_3X_3)(1+a_4X_4)}.
      \end{eqnarray}
      For each $i$, we determine $X_i$,
         \begin{eqnarray}
        & S_i^{-1} =
           (1-a_iX_i)(1+a_iX_i)^{-1}&\nonumber\\
           &Q_{i-}^{-1} Q_{i+} = (a_iX_i+1)(a_iX_i-1)^{-1}&\nonumber\\
         &    Q_{i+} (a_iX_i-1) = Q_{i-}(a_iX_i+1)&\nonumber\\
           &  a(Q_{i+} -Q_{i-}) X_i =  Q_{i+} +Q_{i-} & \nonumber\\
           & a_i \gamma_5 ( (b_i-c_i)  D_w + 2  )\gamma_5 X_i = (b_i+c_i) \gamma_5 D_w .&
     \end{eqnarray}
         This results in,
         \begin{eqnarray}\label{argument}
       a_i X_i = (b_i+c_i) \gamma_5 D_w \frac{1}{2+(b_i-c_i)D}.
         \end{eqnarray}
We can therefore write,
         \begin{eqnarray}
 D^5_{OV}(m) =
    \left[\begin{array}{cccc}
       D^4_{OV}(m) & 0 & 0 & 0  \\
      0 &  1  & 0 & 0  \\
      0 & 0  & 1  & 0\\
      0 & 0  & 0 & 1  \\
    \end{array} \right].
  \end{eqnarray}
     
\section{Computation of the 4D propagator}
\label{sec:appb}

It follows directly from,
  \begin{eqnarray}
    \left[\begin{array}{cccc}
      D_{OV}^4  & 0 & 0 & 0  \\
      0 &  1  & 0 & 0  \\
      0 & 0  & 1   & 0\\
      0 & 0  & 0 & 1    \\
    \end{array} \right]
    \left(\begin{array}{c} x_1 \\ x_2\\x_3\\x_4 \end{array} \right)
    =     \left(\begin{array}{c} b_1 \\ b_2\\b_3\\b_4 \end{array} \right),
  \end{eqnarray}
  or
    \begin{eqnarray}
  D_{OV}^4  x_1 = b_1 ,
    \end{eqnarray}
    that the 4D propagator is equal to $x_1$. We use eq.(\ref{eq:4d-5d}) and find
      \begin{eqnarray}
         F^{-1} L D_{DW}(m) R \vec{x} = \vec{b} ,
      \end{eqnarray}
or
      \begin{eqnarray}
         R_1^{-1}(1) P^{-1} D_{DW}^{-1}(1) D_{DW}(m) P R_1 \vec{x} = \vec{b} .
      \end{eqnarray}
      It follows from
       \begin{eqnarray}
   R_1(1) D^5_{OV} R_1^{-1} \vec{y} =
     \left[\begin{array}{cccc}
         D^4_{OV} & 0 & 0 & 0  \\
       S_2 S_3S_4 (\gamma_5 D^4_{OV} -c_+) & 1  & 0 & 0  \\
       S_3 S_4 (\gamma_5 D^4_{OV} -c_+)  & 0  & 1  & 0\\
       S_4 (\gamma_5 D^4_{OV} -c_+)  & 0  & 0 & 1   \\
    \end{array} \right] \vec{y} = \vec{b},
  \end{eqnarray}
       that $y_1 = x_1$, i.e.\ that the 4D propagator is not affected
       by the transformation with $R_1$.  Hence we can use
    \begin{eqnarray}
   P^{-1} D_{DW}^{-1}(1) D_{DW}(m) P  \vec{y} = \vec{b} ,
    \end{eqnarray}
or
    \begin{eqnarray}\label{tosolve}
      D_{DW}(m) P  \vec{y} =D_{DW}(1) P \vec{b} ,
    \end{eqnarray}
    to determine the 4D propagator $y_1$.

\end{document}